\documentclass[journal]{IEEEtran}
\usepackage{cite}
\usepackage{amsmath}
\usepackage{url}

\usepackage{amsmath} 
\usepackage{amssymb}  
\usepackage{amsthm}

\usepackage[short]{optidef}

\usepackage{color}
\usepackage{psfrag}
\usepackage{pstricks}
\usepackage{pst-plot}
\usepackage{pst-circ}
\usepackage{enumitem}
\usepackage{graphicx}
\graphicspath{{./}}
\DeclareGraphicsExtensions{.eps}

\newcommand{\tab}{Tab.\,}

\newcommand{\cf}{cf.\,}
\newcommand{\ie}{i.e.\,}
\newcommand{\eg}{e.g.\,}
\newcommand{\degc}{^{\circ}\text{C}}

\newcommand{\un}[1]{\underline{#1}}

\newcommand{\R}[1]{\mathbb{R}^{#1}}
\newcommand{\set}[1]{\mathbb{#1}}
\newcommand{\e}{\begin{equation}}
\newcommand{\ee}{\end{equation}}
\newcommand{\trans}{^{\top}} 
\newcommand{\ind}[1]{^{(#1)}}

\newcommand{\ts}[1]{\text{#1}}

\newcommand{\ctrl}{^{\ts{C}}}
\newcommand{\da}{^{\ts{DA}}}
\newcommand{\id}{^{\ts{ID}}}
\newcommand{\reg}{^{\ts{REG}}}
\newcommand{\rsv}{^{\ts{RES}}}
\newcommand{\sfr}{^{\ts{SFR}}}
\newcommand{\sys}{^{\ts{S}}}

\newcommand{\regup}{^{\ts{up}}}
\newcommand{\regdn}{^{\ts{dn}}}
\newcommand{\intTsys}{\mathbb{T}\sys(s)}

\newcommand{\intTc}{\mathbb{T}\ctrl(l)}
\newcommand{\hor}{T^{\ts{H}}}
\newcommand{\ld}{_{\ts{ld}}}
\newcommand{\rf}{_{\ts{ref}}}
\newcommand{\rfi}[1]{_{\ts{ref},#1}}
\newcommand{\tgt}{_{\ts{tgt}}}
\newcommand{\tgti}[1]{_{\ts{tgt}}^{(\ts{#1})}}

\renewcommand{\set}[1]{\mathbb{#1}}

\begin{document}
%
\title{Unlocking the Potential of Flexible Energy Resources to Help Balance the Power Grid}
%
%
%

\author{Fabian~L.~M\"uller,
		Stefan Woerner,
        and~John~Lygeros,~\IEEEmembership{Fellow,~IEEE}
%
\thanks{F. L. M\"uller is with the Automatic Control Laboratory, Swiss Federal Institute of Technology, Zurich, Switzerland, and IBM Research--Zurich, Zurich, Switzerland.
        {\tt\small fmu@zurich.ibm.com}}%
\thanks{S. Woerner is with IBM Research--Zurich, Zurich, Switzerland.
        {\tt\small wor@zurich.ibm.com}}%
\thanks{J. Lygeros is with the Automatic Control Laboratory, Swiss Federal Institute of Technology, Zurich, Switzerland.
        {\tt\small jlygeros@ethz.ch}}%
}

\maketitle

\begin{abstract}
Flexible energy resources can help balance the power grid by providing different types of ancillary services. However, the balancing potential of most types of resources is restricted by physical constraints such as the size of their energy buffer, limits on power-ramp rates, or control delays. Using the example of Secondary Frequency Regulation, this paper shows how the flexibility of various resources can be exploited more efficiently by considering multiple resources with complementary physical properties and controlling them in a coordinated way. To this end, optimal adjustable control policies are computed based on robust optimization. Our problem formulation takes into account power ramp-rate constraints explicitly, and accurately models the different timescales and lead times of the energy and reserve markets. Simulations demonstrate that aggregations of select resources can offer significantly more regulation capacity than the resources could provide individually.

\end{abstract}

\begin{IEEEkeywords}
Aggregation, frequency regulation, robust optimization 
\end{IEEEkeywords}

\section{Introduction}
\label{s:introduction}

\IEEEPARstart{T}{o} guarantee the safe and reliable operation of the electricity grid, the Independent System Operator (ISO) depends on several types of ancillary services. In particular, the Secondary Frequency Regulation (SFR) service is invoked by the ISO to reestablish the nominal grid frequency and to compensate for mismatches of energy schedules on the timescale of seconds to minutes \cite{Swissgrid2017,ENTSO-E2009}. SFR has traditionally been provided by controllable large-scale generating units, such as fossil-fired steam turbines, combustion turbines, nuclear power plants, and hydro power units. With a growing share of renewable intermittent energy resources in the power grid, such as wind and solar, the need for rapid and accurate SFR is increasing. 
%
To support the traditional providers of SFR, various types of electric energy resources have been identified whose flexibility in producing and/or consuming electricity can be used for SFR. In particular, the balancing potential of heating and cooling systems \cite{Vrettos2014a,Qureshi2016,Hao2013b,Lin2013}, plug-in electric vehicles \cite{Vagropoulos2013,Liu2014}, and systems behaving like batteries \cite{Hao2013} has been studied extensively.

The potential of individual systems to provide SFR can be severely limited by their physical constraints. For instance, plug-in electric vehicle (PEV) batteries, domestic heating and cooling systems, capacitors, and flywheels are mainly limited by the size of their energy buffers \cite{Borsche2013,Lin2013}, while steam turbines and industrial cooling systems must adhere to restrictive power-ramp rate limits \cite{Makarov2008}. Moreover, operational constraints, such as the trips of PEVs, run-time constraints, and control delays, can have significant impact on the ability of a system to provide SFR. In addition to its physical and operational constraints, a system offering SFR services must satisfy the terms and conditions dictated by the energy and reserve markets, such as market timescales and lead times, as well as particular requirements on bids, such as time-invariant and symmetric reserve bids \cite{Swissgrid2017}.
%
Approaches to remedy these issues fall into two main groups. The approaches of the first group propose new SFR products that are better suited for the kind of systems listed above, for example by limiting the energy content \cite{Nosair2015a,Lymperopoulos2015,Vrettos2015} or the frequency bandwith of SFR products \cite{Hao2013b,Borsche2014,Fabietti2016}, or by reducing the time period over which SFR capacity must be available \cite{Vrettos2014a}.
%
The second group of approaches relies on aggregating the flexibility of multiple systems and controlling them in a coordinated way. This has proven useful to alleviate the difficulties related to the duration and symmetry requirements of certain SFR markets \cite{Vrettos2014a,Rey2017}. 
The same aggregation techniques can be used to couple energy storage systems with energy markets, allowing the former to adjust their energy schedules via the markets and, thereby, to exploit their flexibility more efficiently \cite{Borsche2013,Qureshi2016}.

In a market setting, each group of systems is managed by an Aggregator (AG) who acts as its representative and trades energy and ancillary services on respective markets with the goal of making an economic profit \cite{Gkatzikis2013,Qureshi2016}. We refer to the problem of making optimal decisions on energy and reserve markets as the \emph{energy and reserve bidding problem}. The problem has been approached by means of Model Predictive Control \cite{Koch2012}, stochastic optimization \cite{Zhang2015a,Lymperopoulos2015}, and robust optimization \cite{Warrington2013,Vrettos2014a,Vrettos2015,Gorecki2017}, among others. The robust formulations rely on affinely adjustable robust control policies, which were first applied to the regulation power context in \cite{Warrington2012,Bertsimas2013}, building on the initial work in \cite{BenTal2004}.

While considering power and energy constraints of individual resources when solving the energy and reserve bidding problem, power ramp-rate constraints are commonly neglected in the literature. This is a major shortcoming because the accurate provision of SFR services can be compromised by insufficient ramping capabilities. The crucial importance of incorporating ramping constraints has already been recognized in the context of grid balancing \cite{Borsche2014} and unit commitment \cite{Morales-Espana2014,Parvania2016}, and has even lead to the adoption of new performance-based schemes for the remuneration of SFR provision by several ISOs \cite{FERC2011}.

In our formulation of the bidding problem, we adopt the robust formulations of \cite{Warrington2012,Zhang2014,Vrettos2014a,Gorecki2017}, and make two main contributions. 
First, building on the mathematical foundations developed in our previous work (for single systems) \cite{Mueller2017b}, we formulate the energy and reserve bidding problem for an aggregation of multiple systems. In contrast to existing approaches, such as \cite{Warrington2012,Zhang2014,Vrettos2014a,Gorecki2017}, we explicitly take into account power-ramp rate constraints and control delays of individual resources. 
Second, we illustrate how aggregating systems with complementary physical properties makes it possible to exploit their SFR potential more efficiently. In particular, our approach allows us to properly model and aggregate the flexibility of systems subject to power ramp-rates and control delays. Simulations show that coupling such systems with fast-responding systems can increase of the amount of available SFR capacity by 18--125\% depending on system parameters.

%

The energy and reserve bidding problem is introduced in Section \ref{s:problemDescription}. Section \ref{s:reqsPowerRef} introduces the adjustable control setup used. The energetic flexibility of individual system subject to power, power ramp-rate, and energy constraints is provided in Section \ref{s:flexDescription}. It is used in Section \ref{s:scheduling} to formulate the aggregate bidding problem as a robust optimization problem. In Section \ref{s:results1}, we present the results from aggregating the flexibility of systems with complementary physical properties. A conclusion is given in Section \ref{s:conclusion}.

\section{Problem Description}
\label{s:problemDescription}

We consider an AG that controls the electricity production or consumption of a group of flexible energy resources, referred to as a \emph{balance group}, over a planning horizon of duration $\hor$. The AG can exploit the collective flexibility of the balance group when trading on energy and reserve markets. 



\subsection{Energy markets}
To satisfy the energy requirements of the systems in its balance group, the AG can trade electric energy prior to delivery. On the day-ahead market, energy is traded once a day for every time interval of duration $T\da$ (usually ${T\da=1}$ h) of the next day with lead time $T\da_{\ts{ld}}$ (${T\da_{\ts{ld}}=13}$ h in Switzerland). We denote by ${e\da\in\R{N\da}}$, ${N\da:=\hor/T\da}$, the final energy schedule resulting from all the day-ahead market trades executed during the planning horizon. To make adjustments to the day-ahead energy schedule during a particular day, the intra-day market allows trading energy continuously for time intervals of durations $T\id$ (usually ${T\id=5-15}$ min) with lead time $T\id_{\ts{ld}}$ (${T\id_{\ts{ld}}=1}$ h in Switzerland). All intra-day energy trades are summarized by ${e\id\in\R{N\id}}$, ${N\id:=\hor/T\id}$. The total energy procurement costs are ${C^E:=c^{\ts{DA}\trans} e\da+c^{\ts{ID}\trans} e\id}$, where ${c\da\in\R{N\da}}$ and ${c\id\in\R{N\id}}$ denote the energy prices, which are unknown at the time of bidding.

The AG is assumed to be \emph{balance-responsible}, \ie, it must ensure that the energy production/consumption of its balance group complies with the net energy schedule resulting from the markets. The AG is held accountable for any mismatches between the actual and planned energy schedule.

\subsection{Reserve markets}
\label{ss:reserveMarkets}
The ISO constantly observes the state of the power grid and, in case of imbalances, takes measures to reestablish the nominal operating state. To be able to compensate fluctuating generation and consumption and other unexpected disturbances, it procures ancillary services on reserve markets ahead of time \cite{ENTSO-E2009}. We focus on Secondary Frequency Regulation \cite{Swissgrid2017,ENTSO-E2009} and assume that \emph{symmetric} reserve capacities ${\gamma_k\in\R{+}_0}$ are traded for all time intervals of duration $T\sys$ in the reserve tendering period $T\sfr$. The approach can be generalized to asymmetric reserves \cite{Vrettos2015}. For simplicity, we will use ${T\sfr=\hor}$ and, thus, ${k=1,\dots,N\sys:=\hor/T\sys}$. If the capacity bid $\gamma_k$ of an AG is accepted, the ISO has the right, but not the obligation, to ask the AG to deviate from the planned power reference by at most $\pm\gamma_k$ units of power at any time during the corresponding time interval ${[(k-1)T\sys,kT\sys)}$. In return for keeping the reserve capacity ${\gamma:=[\gamma_1,\dots,\gamma_{N\sys}]\trans}$ available over the horizon $\hor$, the AG receives the capacity reservation payment ${R\sfr:=c^{\ts{SFR}^\top}\gamma}$, where $c\sfr$ are the reserve capacity prices, which are unknown at the time of bidding \cite{Swissgrid2017}. 
Power deviations due to SFR activation can result in energy deviations from the planned energy schedule of an AG. This so-called up- and down-regulation energy is usually measured over each time interval $T\id$ separately for positive and negative activation, and is denoted by $e\regup$ and $e\regdn$, respectively  \cite{Swissgrid2017}. The AG is remunerated for providing SFR based on the amount of up- and down-regulation energy delivered, \ie, the remuneration is  ${R\reg:=c^{\ts{up}\trans} e\regup - c^{\ts{dn}\trans} e\regdn,}$ where ${c\regup,c\regdn\in\R{N\id}}$ are the corresponding regulation energy prices \cite{Swissgrid2017}.  However, recent regulatory changes in the USA require all ISOs to base their SFR compensation payment not only on the amount of regulation power reserved and regulation energy delivered, but also on the achieved accuracy of up- and down-regulation \cite{FERC2011}. This motivates us to impose strict requirements on SFR performance, \cf\ref{ss:powerTarget}.

\subsection{The bidding problem}
\label{ss:problemDescription}
The energy and reserve bidding problem of the AG is to decide how to trade on the energy markets and how much reserve capacity to offer on the SFR market so as to maximize the total expected profit. These decisions must be made subject to the constraints that the energy needs of the systems in the balance group are satisfied and the offered SFR capacity is kept available over the entire planning horizon. 
The more reserve capacity the AG offers, the more restrictive become the constraints on possible trades in the energy markets. Consequently, offering reserves and trading energy involves a trade-off between the reserve reward $R\sfr$, the potential remuneration of regulation energy $R\reg$, and the energy procurement costs $C^E$.

The bidding problem takes place on different timescales. Trading decisions are made on the timescales $T\sfr$, $T\da$, and $T\id$ corresponding to the reserve market, and the day-ahead and the intra-day energy market, respectively. Online activation of SFR occurs at the faster timescale $T\ctrl$. The dynamics of individual systems are time-discretized on an intermediate timescale $T\sys$ commonly chosen in the range of 5--15 min in the case of SFR. In general, it holds that ${\hor\geq T\sfr\geq T\da\geq T\id\geq T\sys\gg T\ctrl}$. For simplicity, we assume that longer time horizons are integer multiples of shorter ones, and for ${\ast\in\{\ts{SFR},\ts{DA},\ts{ID},\ts{S},\ts{C}\}}$ define ${N^{\ast}:=\hor/T^{\ast}}$, and $\set{T}^{\ast}(k)$ the continuous time interval ${[(k-1)T^{\ast},kT^{\ast})}$. Finally, let ${\set{N}^{k}:=\{1,2,\dots,k\}}$ for ${k\in\set{N}^+}$. 
Different timescales are used in different market regions. All timescales $T^{\ast}$ are used as parameters in our approach making it versatile and applicable to the various market settings. Here we consider the Swiss market setting and use ${\hor=T\rsv=1}$ week, ${T\da=1}$ h, ${T\id=15}$ min, ${T\sys=5}$ min, and ${T\ctrl=1}$ s. 


%
%
%
%
%
%
%
%
%
%

\section{Adjustable Power Reference}
\label{s:reqsPowerRef}

\subsection{Power reference and power target}
\label{ss:powerTarget}
Every system ${j=1,\dots,J}$ in the balance group implements a continuous-time piece-wise affine and continuous power reference which is fully defined by its breakpoints ${p\ind{j}\rf:=[p\ind{j}\rfi{0},\dots,p\ind{j}\rfi{N\sys}]\trans}$, according to
\begin{equation}
	p\ind{j}\rf(t):=p\ind{j}\rfi{s-1}+(p\ind{j}\rfi{s}-p\ind{j}\rfi{s-1})(t-(s-1)T\sys)/T\sys,
	\label{eq:defContPref}
\end{equation}
for ${t\in\intTsys}$, ${s\in\set{N}\sys}$. On the aggregate level, the balance group thus implements the reference ${p\ind{agg}\rf(t) := p\ind{1}\rf(t)+\dots+p\ind{J}\rf(t)}$.
The SFR service is activated by the ISO by sending out an activation signal to all systems that offer SFR for the corresponding time interval. The activation signal ${w:=[w_1,\dots,w_{N\ctrl}]\trans}$ is a discrete-time signal with ${w_l\in\set{W}:=[-1,1]}$, ${l\in\set{N}\ctrl}$. The signal is computed by the ISO and broadcast \emph{sequentially} on the timescale $T\ctrl$ (\eg, ${T\ctrl=1-5}$ s in Continental Europe \cite{ENTSO-E2009}, ${T\ctrl=1}$ s in Switzerland \cite{Lymperopoulos2015}, and ${T\ctrl=4-6}$ s in the USA \cite{FERC2011}) such that $w_l$ becomes available only at time $(l-1)T\ctrl$. We interpret the discrete signal $w$ as the continuous-time piece-wise affine and continuous activation signal
\begin{equation*}
	w(t):= w_{l-1}+(w_l-w_{l-1})(t-(l-1)T\ctrl)/T\ctrl,
\end{equation*}
where ${t\in\intTc}$, ${l\in\set{N}^{C}}$. We assume that an AG offering the SFR capacity $\gamma\ind{agg}\in\R{N\sys}$ is responsible for its balance group to continuously track the target power level
\begin{equation}
	p\tgti{agg}(t,w):=p\ind{agg}\rf(t)+\gamma\ind{agg}_s w(t),
	\label{eq:defAggTargetPower}
\end{equation}
with ${t\in[0,\hor]}$ and ${s\in\set{N}\sys: t\in\intTsys}$. This assumption is more restrictive than current European regulations that only require tracking $w(t)$ within a certain tolerance \cite{TestSC}. However, the accurate provision of SFR is not only desirable from a grid-balancing perspective, but it is already considered in the SFR compensation payment by ISOs in the USA, \cf Section \ref{ss:reserveMarkets}. Inaccurate provision of SFR thus reduces the profit made by the AG and can even lead to disqualification.

The SFR delivered by an AG is the result of a collective effort of all the systems in its balance group, \ie, ${\gamma\ind{agg}:=\gamma\ind{1}+\dots+\gamma\ind{J}}$. Consequently, each system is required to follow its own target power trajectory
\begin{equation}
	p\tgti{j}(t,w):=p\ind{j}\rf(t)+\gamma\ind{j}_s w(t).
	\label{eq:defTargetPower}
\end{equation}
Compared with $p\ind{j}\rf(t)$, which is piece-wise affine on the timescale $T\sys$, the activation $w(t)$ can vary at the higher rate $T\ctrl$. Thus, accurate tracking of \eqref{eq:defTargetPower} puts higher requirements on the ramping capabilities and control delays of a system.

\subsection{Adjustable power reference}
\label{ss:adjustablePower}
To exploit the flexibility of energy resources more efficiently, it has been proposed to adjust the reference power schedules \eqref{eq:defContPref} in response to past activation, \cf \cite{Warrington2012,Bertsimas2013,Vrettos2015,Gorecki2017}, among others. Let ${\tilde{w}\in\set{W}^{N\sys}}$ denote the activation signal $w(t)$ averaged over time intervals of duration $T\sys$, \ie,
\begin{equation*}
	\tilde{w}_s:=\frac{1}{T\sys}\int_{(s-1)T\sys}^{sT\sys} w(t) dt,\ s\in\set{N}\sys.
\end{equation*}
Here we consider power references \eqref{eq:defContPref} whose breakpoints $p\rf\ind{j}$ are affinely adjustable based on past activation, \ie,
\begin{equation}
	p\rf\ind{j}(\tilde{w}) = Q\ind{j}\tilde{w}+q\ind{j},\ j\in\set{N}^J,
	\label{eq:defControllers}
\end{equation}
with parameters ${Q\ind{j}\in\R{(N\sys+1)\times N\sys}}$ and ${q\ind{j}\in\R{N\sys+1}}$. Equation \eqref{eq:defControllers} is a control policy describing the power reference a system should follow depending on a certain SFR activation. The policy must be causal, \ie, it can depend on past activation only. The value $\tilde{w}_s$ becomes known only at time $sT\sys$ and may be used exclusively to adjust future breakpoint values $p\rfi{s+1}\ind{j},\dots,p\rfi{N\sys}\ind{j}$. Thus, the causality of \eqref{eq:defControllers} requires that 
\begin{equation}
	Q\ind{j}_{m,n} = 0\ \forall m\in\set{N}^{\ts{S}+1},\,n\in\set{N}\sys: n\geq m-1.
	\label{eq:reqCausality}
\end{equation}

\subsection{Energy balance}
\label{ss:tradingConstr}
To comply with the energy contracts concluded on the day-ahead and intra-day markets, the balance-responsible AG must ensure that its balance group follows the net power references $p\da$ and $p\id$ resulting from the day-ahead and intra-day energy markets, respectively, \ie
\begin{equation}
	p\rf\ind{agg}(\tilde{w})= p\da + p\id\ \forall \tilde{w}\in\set{W}^{N\sys}.
	\label{eq:balanceEquation}
\end{equation}
Given our choice of affine policies \eqref{eq:defControllers}, the balance equation \eqref{eq:balanceEquation} suggests that the day-ahead and intra-day trading decisions be affine policies of $\tilde{w}$ also, \ie
\begin{align}
\begin{split}
	p\da(\tilde{w}) &= Q\da\tilde{w}+q\da,\\
	p\id(\tilde{w}) &= Q\id\tilde{w}+q\id.
\end{split}
 	\label{eq:powerTradePolicies}
\end{align}
In this case, the balance equation \eqref{eq:balanceEquation} is satisfied if and only if
\begin{align}
\begin{split}
	Q\ind{agg}:=&\sum\limits_{j\in\set{N}^J}Q\ind{j} = Q\da+Q\id,\ \ts{and}\\
	q\ind{agg}:=&\sum\limits_{j\in\set{N}^J}q\ind{j} = q\da+q\id,
\end{split}
\end{align}
with parameters ${Q\da,Q\id\in\R{(N\sys+1)\times N\sys}}$ and ${q\da,q\id\in\R{N\sys+1}}$.
The power policies \eqref{eq:powerTradePolicies} correspond to the energy trading policies
\begin{align}
\begin{split}
	 e\da(\tilde{w}) &= E\da(Q\da\tilde{w}+q\da),\\
	 e\id(\tilde{w}) &= E\id(Q\id\tilde{w}+q\id),
	\label{eq:energyTradePolicies}
\end{split}
\end{align}
where ${E\da\in\R{N\da\times(N\sys+1)}}$ and ${E\id\in\R{N\id\times(N\sys+1)}}$ map power to energy. The AG can use the policies \eqref{eq:energyTradePolicies} to decide on the amount of energy to trade on the corresponding markets based on past SFR activation data. To be applicable, the policies must, however, be causal and take into account the lead times of the different energy markets. These requirements can be incorporated by imposing particular structures onto the matrices $(E\da Q\da)$ and $(E\id Q\id)$. 
\section{Description of flexibility}
\label{s:flexDescription}
The decisions the AG makes on the different energy and reserve markets are restricted by the requirement that the individual target power trajectories \eqref{eq:defTargetPower} must satisfy all the physical constraints of the corresponding system ${j\in\set{N}\sys}$ for all realizations of the unknown activation signal. We consider power, power ramp-rate, and state (\eg energy) constraints. The system index $(j)$ is omitted to simplify the notation.

\subsection{Power constraints}
\label{ss:powerConstr}
The power the system can draw from or feed into the power grid is limited. For all ${t\in\intTsys}$, ${k\in\set{N}\sys}$, we require that
\begin{equation}
	\un{p}_k \leq p\tgt(t,w)\leq \bar{p}_k,\ \forall w\in\set{W}^{N\ctrl}, 
	\label{eq:powerConstrCt}
\end{equation}
where ${\un{p},\bar{p}\in\R{N\sys}}$ denote the piece-wise constant bounds on power. The constraints above are satisfied if for all ${\tilde{w} \in\set{W}^{N\sys}}$ and ${k=0,\dots,N\sys}$ it holds that 
\begin{align}
\begin{split}
	p\rfi{k}(\tilde{w}) + \gamma_{\max\{1,k\}} &\leq \bar{p}_{\max\{1,k\}},\\
	p\rfi{k}(\tilde{w}) - \gamma_{\max\{1,k\}} &\geq \un{p}_{\max\{1,k\}},\\
	p\rfi{k}(\tilde{w}) + \gamma_{\min\{N\sys,k+1\}} &\leq \min\{\bar{p}_k,\bar{p}_{\min\{N\sys,k+1\}}\},\\
	p\rfi{k}(\tilde{w}) - \gamma_{\min\{N\sys,k+1\}} &\geq \max\{\un{p}_k,\un{p}_{\min\{N\sys,k+1\}}\}.
\end{split}
\label{eq:powerConstr}
\end{align}
The above set of constraints comprises an infinite number of inequalities due to its dependency on $\tilde{w}$, and, thus, cannot be included in an optimization problem as is. However, following the approach in \cite{BenTal2004,Bertsimas2011}, and exploiting the linearity of $\eqref{eq:powerConstr}$ in $\tilde{w}$ and the particular structure of $\set{W}$, allows us to derive a set of equivalent constraints. For example, consider the first inequality in \eqref{eq:powerConstr} and denote by $Q_k$ the $k^{\ts{th}}$ row of matrix $Q$. By exploiting
\begin{equation*}
	p\rfi{k}(\tilde{w})=Q_k\tilde{w}+q_k\leq \|Q_k\|_1+q_k, \forall\tilde{w}\in\set{W}^{N\ctrl},
\end{equation*}
the first inequality in $\eqref{eq:powerConstr}$ is equivalent to
\begin{equation*}
	\|Q_k\|_1+q_k+ \gamma_{\max\{1,k\}} \leq \bar{p}_{\max\{1,k\}},
\end{equation*}
which is independent of $\tilde{w}$. The remaining constraints in \eqref{eq:powerConstr} can be reformulated similarly.


\subsection{Power ramp-rate constraints}
\label{ss:rampConstr}	
Limits on the rate at which power can vary over time play an important role, in particular for providing ancillary services with high accuracy. For all ${t\in\intTsys}$, ${k\in\set{N}\sys}$, we require that the rate of change of the target power level be bounded, \ie,
\begin{equation}
	\un{r}_k \leq \frac{\partial}{\partial t}p\tgt(t,w) \leq \bar{r}_k,\ \forall w\in\set{W}^{N\ctrl}, 
	\label{eq:rampConstrCt}
\end{equation}
where ${\un{r},\bar{r}\in\R{N\sys}}$ denote the piece-wise constant power ramp-rate limits. The above inequalities are satisfied if and only if for all ${\tilde{w} \in\set{W}^{N\sys}}$ it holds for ${k=1,\dots,N\sys}$ that
\begin{align}
\begin{split}
	(p\rfi{k}(\tilde{w})-p\rfi{k-1}(\tilde{w}))/T\sys + 2\gamma_k/T\ctrl &\leq \bar{r}_k,\\
	(p\rfi{k}(\tilde{w})-p\rfi{k-1}(\tilde{w}))/T\sys - 2\gamma_k/T\ctrl &\geq \un{r}_k,
	\end{split}
\label{eq:rampConstr1}
\end{align}
and for ${k=1,\dots,N\sys-1}$ that
\begin{align}
\begin{split}
	(p\rfi{k}(\tilde{w})-p\rfi{k-1}(\tilde{w}))/T\sys + (\gamma_k+\gamma_{k+1})/T\ctrl &\leq \bar{r}_k,\\
	(p\rfi{k}(\tilde{w})-p\rfi{k-1}(\tilde{w}))/T\sys - (\gamma_k+\gamma_{k+1})/T\ctrl &\geq \un{r}_k.
		\end{split}
	\label{eq:rampConstr2}
\end{align}
Constraints equivalent to \eqref{eq:rampConstr1} and \eqref{eq:rampConstr2} but independent of $\tilde{w}$ can be derived as shown in Section \ref{ss:powerConstr}.

\subsection{State constraints}
Many types of flexible energy resources owe their flexibility to an energy buffer with limited capacity. Examples of such systems are pumped hydro-power plants, batteries, heating and cooling systems, capacitors, and flywheels. Here we consider systems whose energy buffer level ${x(t)\in\R{}}$ is governed by the linear time-invariant dynamics
\begin{equation}
	\frac{\partial}{\partial t}x(t,w)=ax(t,w)+bu_s+cp\tgt(t,w),
	\label{eq:stateDynamics}
\end{equation}
with ${t\in\intTsys}$, ${s\in\set{N}\sys}$. Exogenous (uncontrollable) inputs, such as the weather conditions in the case of heating systems, or trips in the case of electric-vehicle batteries, are summarized by ${u:=[u_1,\dots,u_{N\sys}]\trans}$ and are assumed to be piece-wise constant and known. Because the bidding problem has to be solved ahead of time, the initial state $x(0)$ is not known precisely, but is known to lie in the interval ${[x_{0,\min},x_{0,\max}]}$. The dynamics \eqref{eq:stateDynamics} are characterized by the parameters ${\{a,b,c\}}$, where ${a\leq 0}$ can be interpreted as the self-dissipation rate of the energy buffer, and $b,c$ determine the conversion efficiencies of exogenous inputs and electric energy into buffered energy, respectively. The behavior of an ideal battery, for instance, is modeled by ${a=b=0}$ and ${c=1}$.

We consider the piece-wise constant state constraints
\begin{equation}
	\un{x}_{s}\leq x(t,w) \leq\bar{x}_{s},\ \forall w\in\set{W}^{N\ctrl},
	\label{eq:stateConstrCt}
\end{equation}
for all ${t\in\intTsys}$, ${s=\set{N}\sys}$, ${x(0)\in[x_{0,\min},x_{0,\max}]}$, where ${\un{x},\bar{x}\in\R{N\sys}}$ denote the physical limits of the system state. The fact that the target power $p\tgt$ is piece-wise affine on the timescale $T\ctrl$ makes the robust reformulation of \eqref{eq:stateConstrCt} nontrivial. Due to space limitations, the interested reader is referred to the detailed discussion in \cite{Mueller2017b}, where the robust counterparts of \eqref{eq:stateConstrCt} are derived and shown to be linear in the trading policy parameters $\{Q\ind{j},q\ind{j}\}$ of an individual resources.

\subsection{Description of flexibility}

The robust counterparts of the power, ramp-rate, and state constraints \eqref{eq:powerConstrCt}, \eqref{eq:rampConstrCt}, and \eqref{eq:stateConstrCt}, respectively, are linear in the decision variables ${\zeta\ind{j}:=\{Q\ind{j},q\ind{j},\gamma\ind{j}\}}$, \cf\cite{Mueller2017b}. Thus, they define a convex polytope ${\set{P}(\Phi\ind{j})}$ parameterized by the system constraints and dynamics parameters $\Phi\ind{j}:=\{\un{p}\ind{j},\bar{p}\ind{j},\un{r}\ind{j},\bar{r}\ind{j},\un{x}\ind{j},\bar{x}\ind{j},x\ind{j}_{0,\min},x\ind{j}_{0,\max},a\ind{j},b\ind{j},c\ind{j}\}$. The polytope ${\set{P}(\Phi\ind{j})}$ can serve as a concise and convenient description of the flexibility system $j$ offers with regard to trading energy and providing SFR services.
\section{The Bidding Problem}
\label{s:scheduling}

\subsection{Problem formulation}
The profit made by the aggregator is
\begin{equation*}
J(\Omega,\Psi) :=\ R\rsv(\Omega)+R\reg(\Omega,\Psi)-C^E(\Omega,\Psi),
\end{equation*}
where we have summarized all decision variables in ${\Omega:=\{\zeta\ind{1},\dots,\zeta\ind{J},Q\da,q\da,Q\id,q\id\}}$, and all the uncertain parameters in ${\Psi:=\{w,c\da,c\id,c\rsv,c^{\ts{up}},c^{\ts{dn}}\}}$. The energy and reserve bidding problem consists of finding the power reference policies \eqref{eq:defControllers}, the trading policies \eqref{eq:powerTradePolicies}, and the SFR capacities $\gamma\ind{j}$ that maximize the expected profit while satisfying all the system and market constraints. The problem can be written as the linear program
\begin{maxi!}[2]
	{\Omega}{\operatornamewithlimits{\mathbf{E}}\limits_{\Psi}[ J(\Omega,\Psi)]}
	{\label{eq:schedulingProbEco}}{}
	\addConstraint{\zeta\ind{j}}{\in\set{P}(\Phi\ind{j}),\ j\in\set{N}^J \label{eq:feasibilityConstr}}
	\addConstraint{\sum_{j\in\set{N}^J}Q\ind{j}}{ = Q\da + Q\id \label{eq:QConstr}}
	\addConstraint{\sum_{j\in\set{N}^J}q\ind{j}}{ = q\da + q\id \label{eq:qConstr}}
	\addConstraint{\ts{Causality of \eqref{eq:defControllers} via \eqref{eq:reqCausality}}\label{eq:causalityConstr}}{}
	\addConstraint{\ts{Market constraints.}\label{eq:marketConstr}}{}
\end{maxi!}
The constraints \eqref{eq:feasibilityConstr} ensure the feasibility of the target power trajectory \eqref{eq:defTargetPower} with respect to the constraints of the inidividual system ${j\in\set{N}^J}$. The constraints \eqref{eq:QConstr} and \eqref{eq:qConstr} guarantee that the AG complies with all energy market trades. The power reference policies must be causal \eqref{eq:causalityConstr}, and respect all market constraints \eqref{eq:marketConstr}, such as market lead times or the requirement of time-invariant reserve bids \cite{Swissgrid2017}.

\subsection{Complexity of the bidding problem}
The number of decision variables and constraints of problem \eqref{eq:schedulingProbEco} is determined by various factors, such as the length of the planning horizon, the system discretization time, the different market timescales, and also the structure of the $Q$-matrices that determine the power reference and trading policies \eqref{eq:defControllers} and \eqref{eq:powerTradePolicies}, respectively. For instance, the problem size can be reduced significantly, if the power reference policy \eqref{eq:defControllers} considers only the ${r\geq 1}$ most recent measurements of the SFR activation signal instead of its entire history. That is, in addition to \eqref{eq:reqCausality} we require that
\begin{equation*}
	Q\ind{j}_{m,n} = 0\ \forall m\in\set{N}^{\ts{S}+1},\,n\in\set{N}\sys: m>n+r+2,
\end{equation*}
which imposes a band structure onto $Q\ind{j}$. The effects of different policies on the solution of problem \eqref{eq:schedulingProbEco} and the required solving time have been investigated in \cite{Warrington2012,Mueller2017b}.

Solving problem \eqref{eq:schedulingProbEco} for aggregations of a large number of systems can be inefficient. However, the particular structure of the problem, that is its decomposable objective and the linear coupling constraints \eqref{eq:QConstr} and \eqref{eq:qConstr}, allow us to apply decentralized optimization methods, such as ADMM, \cf\cite{Rey2017,Bitlislioglu2017} for more details.

%
\section{Synergy effects of aggregation}
\label{s:results1}


Aggregating the flexibility of multiple energy resources into balance groups has been shown to alleviate some of the challenges faced by individual systems, such as limits on the minimum bid size and the requirement of time-invariant SFR bids \cite{Vrettos2014a,Rey2017}. Here, we show that aggregating the flexibility of systems with complementary physical properties can significantly increase the amount of SFR capacity that can be offered by the aggregation.

For all the simulations discussed below we consider a planning horizon of duration ${\hor=1}$ d, an intra-day market operating on the timescale ${T\id=15}$ min with a lead time of ${T\id\ld=1}$ h, and an SFR market that accepts only \emph{symmetric and time-invariant} capacity bids \cite{Swissgrid2017} for a tendering period of ${T\rsv=1}$ d. The dynamics of the systems are time-discretized on the timescale ${T\sys=5}$ min, and the SFR service is activated on the timescale ${T\ctrl=10}$ s.


\subsection{Battery and freezer with delay}
\label{ss:synergy1}
We consider two flexible systems, an ideal battery (B), and an industrial freezer warehouse (F).
%
The battery is modeled as an ideal energy buffer whose state dynamics are governed by  \eqref{eq:stateDynamics} with parameters ${a\ind{B}=b\ind{B}=0}$ and ${c\ind{B}=1}$.  The ramping capabilities of the battery are assumed to be fast enough so that the power-ramp rate constraints \eqref{eq:rampConstrCt} can be neglected. However, the battery is subjected to the power constraints \eqref{eq:powerConstrCt} with symmetric limits ${-\un{p}\ind{B}=\bar{p}\ind{B}}$, and state constraints \eqref{eq:stateConstrCt} with limits ${\un{x}\ind{B}=0,\,\bar{x}\ind{B}>0}$ and ${x_{0,\min}\ind{B}=x_{0,\max}\ind{B}=\bar{x}\ind{B}/2}$. The maximum SFR capacity the battery can offer is
\begin{multline}
	\gamma^{(B)}_{\max} = \min\{|\bar{p}\ind{B}-\un{p}\ind{B}|/2,(\bar{x}\ind{B}-x_{0,\max}\ind{B})/T\sfr, \dots\\
	(x_{0,\min}\ind{B}-\un{x}\ind{B})/T\sfr\}.
	\label{eq:batCapAlone}
\end{multline}
For many types of batteries, $\gamma^{(B)}_{\max}$ is severly limited by the state constraints \eqref{eq:stateConstrCt} because their energy capacity is usually small compared to the capacity that would be required to offer all of their rated power as SFR capacity. Consider for example the battery of a \textit{Tesla Model-S}  with parameters ${\bar{p}\ind{B}=17.2}$ kW and ${\bar{x}\ind{B}=100}$ kWh \cite{TeslaModelS} bidding into an SFR market. According to \eqref{eq:batCapAlone}, the battery can provide ${\gamma^{(B)}_{\max}=2.08}$ kW which corresponds to 12.1\% of $\bar{p}\ind{B}$. 

The freezer is modeled as a single energy buffer whose state evolves according to \eqref{eq:stateDynamics}. It must respect the power, power ramp-rate, and state constraints \eqref{eq:powerConstrCt}, \eqref{eq:rampConstrCt}, and \eqref{eq:stateConstrCt}, respectively. The parameters of the freezer are similar to the ones of the freezer warehouse described in \cite{Mueller2013} and are given in \tab\ref{t:freezerParams}. The freezer consumes electric energy to keep its indoor air temperature $\theta^{\ts{in}}$ within the admissible range ${[\theta^{\ts{in}}_{\min},\,\theta^{\ts{in}}_{\max}]}$. The time required to cool the freezer from $\theta^{\ts{in}}_{\max}$ down to $\theta^{\ts{in}}_{\min}$ at maximum cooling power $\bar{p}\ind{F}$ is assumed to be 6 h. That is, the freezer has an energy storage capacity of ${\bar{x}\ind{F}=6\cdot\bar{p}\ind{F}=1.8}$ MW. Given the constant outdoor air temperature ${\theta^{\ts{out}}=5\degc}$, we assume that without cooling the indoor temperature rises from $\theta^{\ts{in}}_{\min}$ to $\theta^{\ts{in}}_{\max}$ in ${T^{\ts{dis}}=10}$ h. The parameters of the freezer state dynamics \eqref{eq:stateDynamics} are
\begin{align*}
	a\ind{F} &= \frac{1}{T^{\ts{dis}}}\ln\left(\frac{\theta^{\ts{in}}_{\max}-\theta^{\ts{in}}_{\min}}{\theta^{\ts{in}}_{\min}-\theta^{\ts{out}}}\right),\\
	b\ind{F} &= a\ind{F}\bar{x}\ind{F}/(\theta^{\ts{in}}_{\min}-\theta^{\ts{in}}_{\max}),\\
	c\ind{F} &= 1.
\end{align*} 
Further, we assume that the power consumption of the freezer's cooling system can be set to any value in the continuous range $[\un{p}\ind{F},\bar{p}\ind{F}]$. However, power set point changes are not implemented immediately but only after a delay ${\delta\ind{F}>T\ctrl}$. This delay makes it impossible for the freezer to track an SFR target power trajectory \eqref{eq:defTargetPower} because it cannot react on the time scale $T\ctrl$. Consequently, the freezer cannot offer any SFR, \ie ${\gamma^{(F)}_{\max}=0}$ kW. The control delay $\delta\ind{F}$ is incorporated explicitly in the bidding problem by forcing additional elements of $Q\ind{F}$ to zero:
\begin{equation}
	Q\ind{F}_{m,n} = 0\ \forall m\in\set{N}^{\ts{S}+1},\,n\in\set{N}\sys: n\geq m-1-\left\lceil\frac{\delta\ind{F}}{T\sys}\right\rceil.
	\label{eq:constrQF}
\end{equation}

\begin{table}[!t]
\renewcommand{\arraystretch}{1.3} \caption{Time-invariant freezer parameters} \label{t:freezerParams} \centering
\begin{tabular}{crl} \hline
\bfseries Symbol & \bfseries Value & \bfseries Description\\ \hline
$\un{p}\ind{F},\,\bar{p}\ind{F} $ & 0, 300 kW & Min., max. power consumption\\ 
$\un{r}\ind{F},\,\bar{r}\ind{F}$ & -100, 100 kW/min & Min., max. power ramp-rate\\ 
$\un{x}\ind{F},\,\bar{x}\ind{F}$ & 0, 1.8 MWh & Min., max. state\\
${\un{x}\ind{F}_{\min}=\bar{x}\ind{F}_{\max}}$ & 0.9 MWh & Limits on initial state\\
$\theta^{\ts{in}}_{\min},\,\theta^{\ts{in}}_{\max}$ & -29, -27 $\degc$ & Admissible indoor temp. range\\
$\theta^{\ts{out}}$ & 5 $\degc$ & Constant outdoor air temp.\\
$a\ind{F}$ & -0.0061 h$^{-1}$ & Self-dissipation rate\\
$b\ind{F}$ & 5.4562 kW/$\degc$ & Heat flux to outside\\
$c\ind{F}$ & 1  & Efficiency of cooling system\\
\hline
\end{tabular} 
\end{table}

Individual systems can adjust their power trajectories only by trading energy on day-ahead or intra-day markets. However, trading energy on short notice in response to a biased SFR activation can be expensive. In contrast, the systems in a balance group
can make bilateral adjustments to their power references while keeping the aggregate power reference unchanged, \ie, without the need for trading energy on markets.
To illustrate the beneficial effect of aggregation, we compute the maximum time-invariant and symmetric SFR capacity ${\gamma^{(\ts{agg})}_{\max}\in\R{+}_0}$ that can be offered by the aggregation of the freezer and the battery by solving the slightly modified bidding problem:
\begin{maxi!}[2]
	{\Omega}{\min\limits_{k\in\set{N}\sys}\{\gamma\ind{B}_k+\gamma\ind{F}_k\}}
	{\label{eq:freezerBatteryOpti}}
	{\gamma^{(\ts{agg})}_{\max}:=}
	\addConstraint{\zeta\ind{j}\in\set{P}(\Phi\ind{j}),\ j\in\{B,F\}}{}
	\addConstraint{Q\ind{B}+Q\ind{F}= Q\da+Q\id}{\label{eq:bfBalConstr1}}
	\addConstraint{q\ind{B}+q\ind{F}= q\da+q\id}{\label{eq:bfBalConstr2}}
	\addConstraint{Q\da=Q\id=0,\, q\id=0}{\label{eq:noMarketConstr}}
	\addConstraint{\ts{Causality of \eqref{eq:defControllers} via \eqref{eq:reqCausality}}}{}
	\addConstraint{\ts{Control delay constraint \eqref{eq:constrQF}.}}{\label{eq:delayConstr}}
\end{maxi!}
By adding the constraints \eqref{eq:noMarketConstr}, we model the case where the \emph{aggregate} power reference \eqref{eq:balanceEquation} is fixed and cannot be adjusted via trades on energy markets. However, the individual power references $p\rf\ind{B}$ and $p\rf\ind{F}$ are still adjustable according to \eqref{eq:defControllers} as long as all adjustments cancel out on the aggregate level. Any market constraints \eqref{eq:marketConstr} can be omitted. The control delay of the freezer is taken into account by  \eqref{eq:delayConstr}. 

Solving \eqref{eq:freezerBatteryOpti} using the \textit{Tesla Model-S} battery parameters yields ${\gamma\ind{agg}_{\max}=\gamma^{(B)\ast}=9.61}$ kW, which is 55.9\% of $\bar{p}\ind{B}$ and represents an increase by a factor of 4.6 compared to the amount of SFR the battery can provide individually. 

\begin{figure}[tb]
\setlength{\abovecaptionskip}{-11mm}
\setlength{\belowcaptionskip}{-10mm}
\centering
\includegraphics[width=0.5\textwidth]{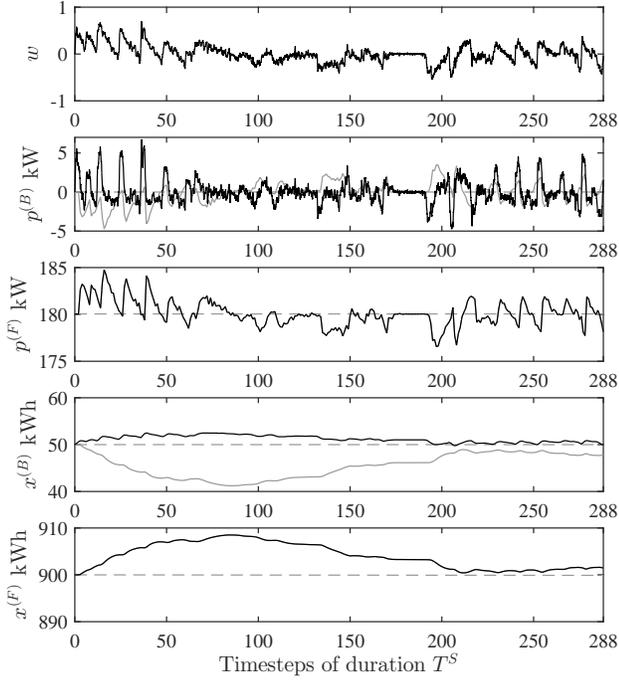}
\caption{Power and state trajectories of the battery and the freezer in response to a given activation signal $w$. Trajectories in case of no activation (${w=0}$) are drawn as dashed grey lines. The reference and target trajectories for the given activation are shown as solid gray and solid black lines, respectively. The reference and target trajectories for the freezer coincide because ${\gamma^{(F)\ast}=0}$.}
\label{fig:freezerBatteryTrajecs}
\end{figure}

Optimal power reference policy parameters are
\begin{align*}
q^{(B)\ast}_k &= 0,\ q^{(F)\ast}_k = q^{\ts{DA}\ast}_k = 180.05,\ k\in\set{N}\sys,\\
Q^{(F)\ast}_{m,n} &= \begin{cases}
	7.59 \mbox{ for } m\in\set{N}^{\ts{S}+1},n\in\set{N}\sys: m=n+3,\\
	0 \mbox{ for } m\in\set{N}^{\ts{S}+1},n\in\set{N}\sys: m\neq n+3,\\
\end{cases}\\
Q^{(B)\ast} &= -Q^{(F)\ast}.
\end{align*}
Figure \ref{fig:freezerBatteryTrajecs} illustrates how the battery and the freezer adjust their power references in response to SFR activation. 
The first plot shows a real activation signal $w$ which was downsampled from 1 s to 10 s resolution over to course of 1 day. The signal is biased toward positive values, in particular during the first half of the planning horizon.
The second plot depicts the power reference $p\rf\ind{B}$ of the battery for the case where no SFR is activated (horizontal dashed grey line at 0 kW), and for the particular activation shown in the first plot (solid grey line). The solid black line denotes the target power level \eqref{eq:defAggTargetPower}. 
The power trajectories of the freezer are provided in the third plot. Because $\gamma^{(F)\ast}=0$, the power reference and power target trajectories of the freezer coincide (solid black line).
Note that ${p\rf\ind{F}=-p\rf\ind{B}}$ as required by the balance constraints \eqref{eq:bfBalConstr1} and \eqref{eq:bfBalConstr2}. 
The evolution of the battery and freezer states $x\ind{B}$ and $x\ind{F}$ are shown in the two bottom plots. In the case of no activation, the states remain constant at their initial values (dashed grey lines). However, for the activation given, both the battery and the freezer must adjust their power references (solid grey line for $x\ind{B}$ and solid black line for $x\ind{F}$). In this way, the battery is able to keep its actual state-of-charge (solid black line) within its feasible range despite a biased activation signal. The energy content of the SFR activation is absorbed by the freezer, whose reference and target state trajectories coincide (solid black line). By acting as a responsive energy buffer, the freezer contributes to providing SFR on the aggregate level even though it cannot offer any SFR on its own. 

%

\begin{figure}[tb]
\centering
\includegraphics[width=0.5\textwidth]{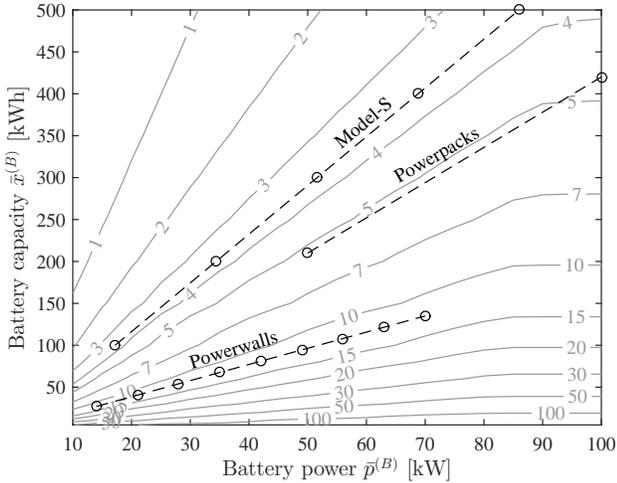}
\caption{Values of the synergy factor $\sigma\sfr$ for different choices of the battery parameters $\bar{p}\ind{B},\,\bar{x}\ind{B}$ for the aggregation of batteries and a freezer.}
\label{fig:synFactor}
\end{figure}

To quantify the benefit of aggregating multiple systems, we define the \emph{SFR synergy factor} as
\begin{equation*}
	\sigma\sfr:=\gamma^{(\ts{agg})}_{\max}\Big/\sum\limits_{j\in\set{N}^J}\gamma^{(j)}_{\max}-1\,\in\R{+}_0.\
\end{equation*}
The synergy factor can take nonnegative values and measures the relative increase of available SFR capacity due to aggregation. A value ${\sigma\sfr>0}$ indicates that aggregation results in a synergy effect and is able to free up SFR capacity that the systems cannot offer individually. Figure \ref{fig:synFactor} provides the $\sigma\sfr$-values for the aggregation of a battery and a freezer for different choices of the battery parameters. The circles indicate the battery parameters corresponding to aggregations of 1--5 \textit{Tesla Model-S} batteries \cite{TeslaModelS}, 1--2 \textit{Tesla Powerpack} batteries \cite{TeslaPowerpack}, and 2-10 \textit{Tesla Powerwall} batteries \cite{Powerwall2016}. The battery parameters and the corresponding individual and aggregate maximum SFR capacities and synergy factors are summarized in \tab\ref{t:synFactor}. The results show that aggregating systems with complementary physical properties -- a fast-responding but energy-constrained battery and a delayed-responding freezer with a comparatively large energy buffer -- can unlock a significant amount of SFR capacity.


\begin{table}[!t]
\renewcommand{\arraystretch}{1.3} \caption{Battery-Freezer Aggregation} \label{t:synFactor} \centering
\begin{tabular}{lcccccc} \hline
Battery type & \# & $\bar{p}\ind{B}$ & $\bar{x}\ind{B}$ & $\gamma^{(\ts{B})}_{\max}$ & $\gamma^{(\ts{agg})}_{\max}$ & $\sigma\sfr$\\ 
     &    & kW & kWh & kW & kW\\ \hline \hline
Model-S   & 1  & 17.2 & 100 & 2.08 & 9.61 & 3.61\\
          & 5  & 86   & 500 & 10.42 & 48.04 & 3.61\\ \hline
Powerpack & 1  & 50   & 210 & 4.38 & 27.09 & 5.19\\
          & 2  & 100  & 420 & 8.75 & 49.47 & 4.65\\ \hline
Powerwall & 2  & 14   & 27 & 0.56 & 7.25 & 11.90\\
          & 10 & 70   & 135 & 2.81& 36.26 & 11.90\\
\hline
\end{tabular} 
\end{table}

\subsection{Battery and turbine with ramp-rate constraints}
\label{ss:synergy2}

Synergy effects similar to the ones discussed above can be observed when coupling a battery with a system whose ability to offer SFR is restricted by power ramp-rate limits. Consider a steam turbine with the time-invariant power limits ${\un{p}\ind{ST}=0}$ MW and ${\bar{p}\ind{ST}=250}$ MW, similar to the \emph{Siemens SST-3000} turbine \cite{Siemens2017}. We assume that the turbine can be operated at full power over the entire planning horizon, \ie the state constraints \eqref{eq:stateConstrCt} are omitted. In contrast to the freezer discussed in Section \ref{ss:synergy1}, the turbine is assumed to be running at an operating point where it can react to power set point changes without delays. However, the rate at which the turbine power can be varied is limited by ${-\un{r}\ind{ST}=\bar{r}\ind{ST}=4.5}$ MW/min \cite{Makarov2008}. These ramp-rates determine the maximum SFR capacity the turbine can offer: ${\gamma\ind{ST}_{\max}=375}$ kW, which corresponds to 0.15\% of its rated power $\bar{p}\ind{ST}$.

We solve the bidding problem \eqref{eq:freezerBatteryOpti} without the control delay constraints \eqref{eq:delayConstr} for different choices of battery parameters $(\bar{p}\ind{B},\bar{x}\ind{B})$. The resulting aggregate SFR capacities and corresponding synergy factors are  given in \tab\ref{t:synFactorSTB}. The results show that the aggregation of a steam turbine and a battery can offer significantly more SFR than the two systems could provide individually. The two systems complement each other: if needed, the turbine can exchange energy with the battery via adjustments of the reference schedules \eqref{eq:defControllers} to keep the state-of-charge of the battery feasible. The battery, on the other hand, takes over the high-frequency component of the activation signal and thereby reduces the thermal stress and operational costs of the turbine. 

\begin{table}[!t]

\renewcommand{\arraystretch}{1.3} \caption{Battery-Steam Turbine Aggregation} \label{t:synFactorSTB} \centering
\begin{tabular}{lcccccc} \hline
Battery type & \# & $\bar{p}\ind{B}$ & $\bar{x}\ind{B}$ & $\gamma^{(\ts{B})}_{\max}$ & $\gamma^{(\ts{agg})}_{\max}$ & $\sigma\sfr$\\ 
     &    & kW & MWh & kW & kW\\ \hline \hline
Model-S   & 10  & 172 & 1 & 20.83 & 468.70 & 0.18\\
          & 50  & 860   & 5 & 104.17 & 843.50 & 0.76\\
          & 100 & 1720 & 10 & 208.33 & 1312.00 & 1.25\\ \hline
Powerpack & 5  & 250   & 1.05 & 21.88 & 506.84 & 0.28\\
          & 10  & 500  & 2.1 & 43.75 & 638.68 & 0.53\\
          & 20 & 1000 & 4.2 & 87.50 & 902.35 & 0.95\\ \hline
Powerwall & 50  &  350  & 0.675 & 14.06 & 551.00 & 0.42\\
          & 100 &  700  & 1.35 & 28.13 & 726.99 & 0.80\\
\hline
\end{tabular} 
\end{table}

\section{Conclusion}
\label{s:conclusion}

We considered the energy and reserve bidding problem of an aggregator that is trading energy and offering SFR on behalf of a group of flexible systems. The problem was formulated as a robust optimization problem that computes optimal affine decision policies. The problem formulation explicitly takes into account power-ramp rates and control delays of the systems, and all the different timescales involved in the bidding problem. It was shown that an aggregation of systems with complementary physical properties can offer significantly more regulation reserves compared with the amount of reserves available from individual systems.

%
\IEEEpeerreviewmaketitle

\section*{Acknowledgment}
The authors gratefully acknowledge the fruitful discussions with Bernhard Jansen.


\ifCLASSOPTIONcaptionsoff
  \newpage
\fi

\bibliographystyle{IEEEtran}
\bibliography{../../../Papers/BibTeX/library}

\enlargethispage{-5in}

\end{document}